\documentclass{article}
\usepackage[T1]{fontenc} 
\usepackage[utf8]{inputenc} 

\def\authorname{S. Kim, H. Lee, S. Park, J. Lee, and K. Choi}

\def\papertitle{Deep Composer Classification Using Symbolic Representation}

\usepackage{ismir-lbd}
\usepackage{xcolor}
\usepackage{hyperref}
\usepackage{pgfplots}

\usepackage{lineno}

\title{\papertitle}

\twoauthors
  {Sunghyeon Kim$^*$, Hyeyoon Lee$^*$, Sunjong Park$^*$, Jinho Lee} {Yonsei University  \\ \tt \{hahala25, hylee817, ryan0507, leejinho\}\\\tt @yonsei.ac.kr}
  {Keunwoo Choi} {ByteDance AI Lab \\ \tt keunwoo.choi\\ \tt @bytedance.com}

\begin{document}

\maketitle
\begin{abstract}
In this study, we train deep neural networks to classify composer on a symbolic domain. The model takes a two-channel two-dimensional input, i.e., onset and note activations of time-pitch representation, which is converted from MIDI recordings and performs a single-label classification.
On the experiments conducted on MAESTRO dataset, we report an F1 value of 0.8333 for the classification of 13~classical composers.

\end{abstract}
\section{Introduction and Proposed System}\label{sec:introduction}
Symbolic representation can be used as input representation to allow the model to be independent to timbre and acoustic recording environment and focus on note-related aspects such as pitch and duration of notes. For example, n-grams of symbolic musical events were used as an input of a composer style classification model in \cite{AoA629}. 





\begin{figure}
 \centerline{\mbox{
 \includegraphics[width=1.0\columnwidth, clip]{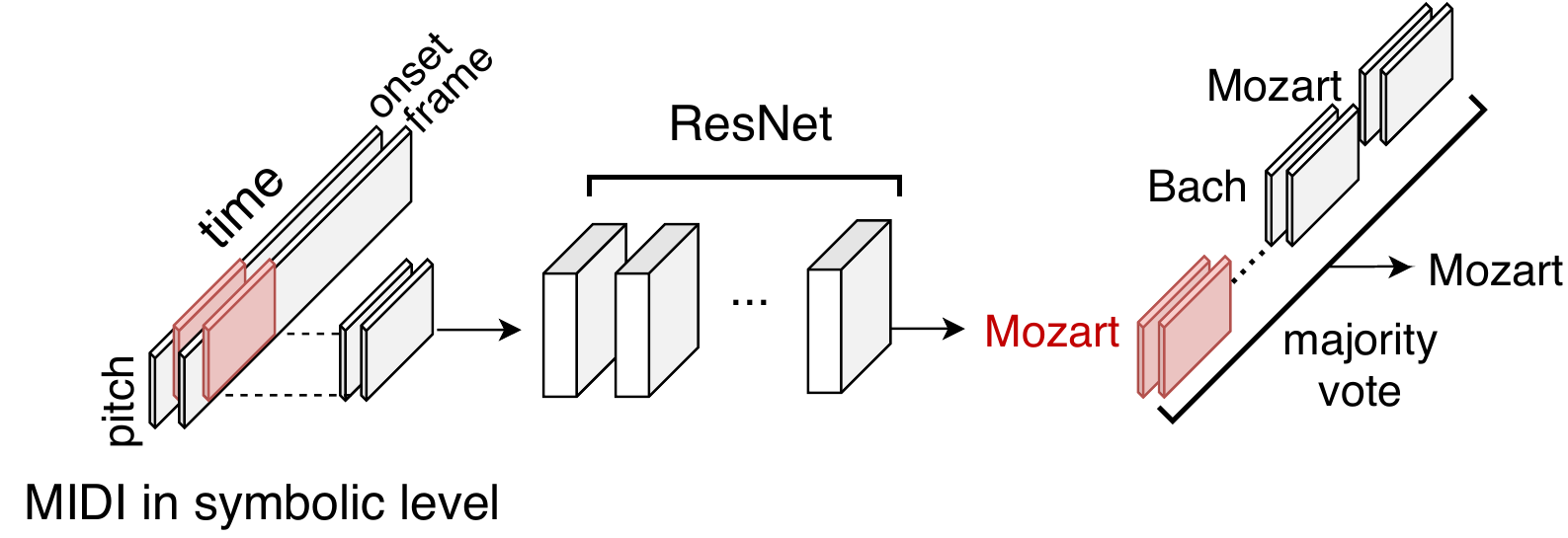}}}
 \caption{A MIDI recording is converted to onset/frame channels which is segmented into the input size of the model. During inference, the final classification of a piece is decided by major voting of segment-wise prediction.
  }
 \label{fig: Process Diagram}
\end{figure}

In the proposed system, as illustrated in Figure \ref{fig: Process Diagram}, we use a ResNet to classify composer from a segment of input music and the final decision is made by majority voting \cite{he2016deep}. For the input, we adopt a symbolic representation that was originally used in piano transcription task \cite{DBLP:journals/corr/abs-1710-11153}. In detail, the midi note events of a recording are converted to onset and frame channels. Each channel is a piano roll-like 2D array and in a shape of (time,~pitch), where a bin represents 0.05~second and 1~semitone, respectively. A bin of the onset channel is binarized where \texttt{1} indicates a note onset. In the frame channel, each bin represents an activation of a note by its recorded midi velocity (0-127).

By designing the proposed system, we expect that the ResNet would learn some specific patterns that could indicate musical styles of composers. This can be done by the network learning spatial features such as pitch interval tendency in the input (e.g. chord and voicing). We also hypothesize that the network would learn some useful velocity patterns from the frame channel such as variations of velocity in a chord or its local/global dynamic range.

\section{Experiments and Discussion}
\subsection{Overview}\label{subsec:data preprocessing}

We used MIDI-only archive of the MAESTRO dataset v2.0.0 \cite{hawthorne2018enabling}. It provides over 200~hours of classical MIDI performances along with metadata such as composer, title, and duration. We excluded pieces with multiple composers to remove any ambiguity 
and selected composers with more than 16~pieces. These steps left 505~pieces by 13~composers in our experiment. Data are split into 7:3 train and test sets (347 and 158 pieces respectively) using stratified sampling with respect to the composer label.

In the dataset, each midi recording corresponds to a piece. To deal with the variable durations, only a randomly selected segment of the whole recording is fed to the model. The duration of a segment is set to 20~second (400 bins). For both training and inference of the model, we use 90~segments that are uniformly sampled over time.\footnote{All the details of the experiment including model and dataset are released at \url{https://github.com/KimSSung/Deep-Composer-Classification}.}

\begin{table}
\footnotesize
 \begin{center}

 \begin{tabular}{@{}c|c||c|c@{}}
  \textbf{Composer (abb.)}\tiny & \textbf{Pieces} & \textbf{Composer (abb.)} & \textbf{Pieces} \\
  \hline
  \hline 
  F. Chopin (Chop) & 64 & W. A. Mozart (Moza) & 29 \\
  \hline
  J. S. Bach (Bach) & 62 & D. Scarlatti (Scar) & 25 \\
  \hline
  L. V. Beethoven (Beet) & 62 & J. Haydn (Hayd) & 20 \\
  \hline
  F. Liszt (Lisz) & 60 & A. Scriabin (Scri) & 19 \\
  \hline
  F. Schubert (F.Sch) & 58 & R. Schumann (R.Sch) & 18 \\
  \hline
  C. Debussy (Debu) & 37 & J. Brahms (Brah) & 17 \\
  \hline
  S. Rachmaninoff (Rach) & 34 &  & \\ 
  
 \end{tabular}
\end{center}
 \caption{List of remaining data classified by its canonical composer: Pieces are the count of unique canonical titles owned by each composer.}
 \label{tab:example}
\end{table}

\begin{figure}
\small
 \centerline{\mbox{
 \includegraphics[width=1.0\columnwidth, trim={0.5cm 0.3cm 1.2cm 2.5cm},clip]{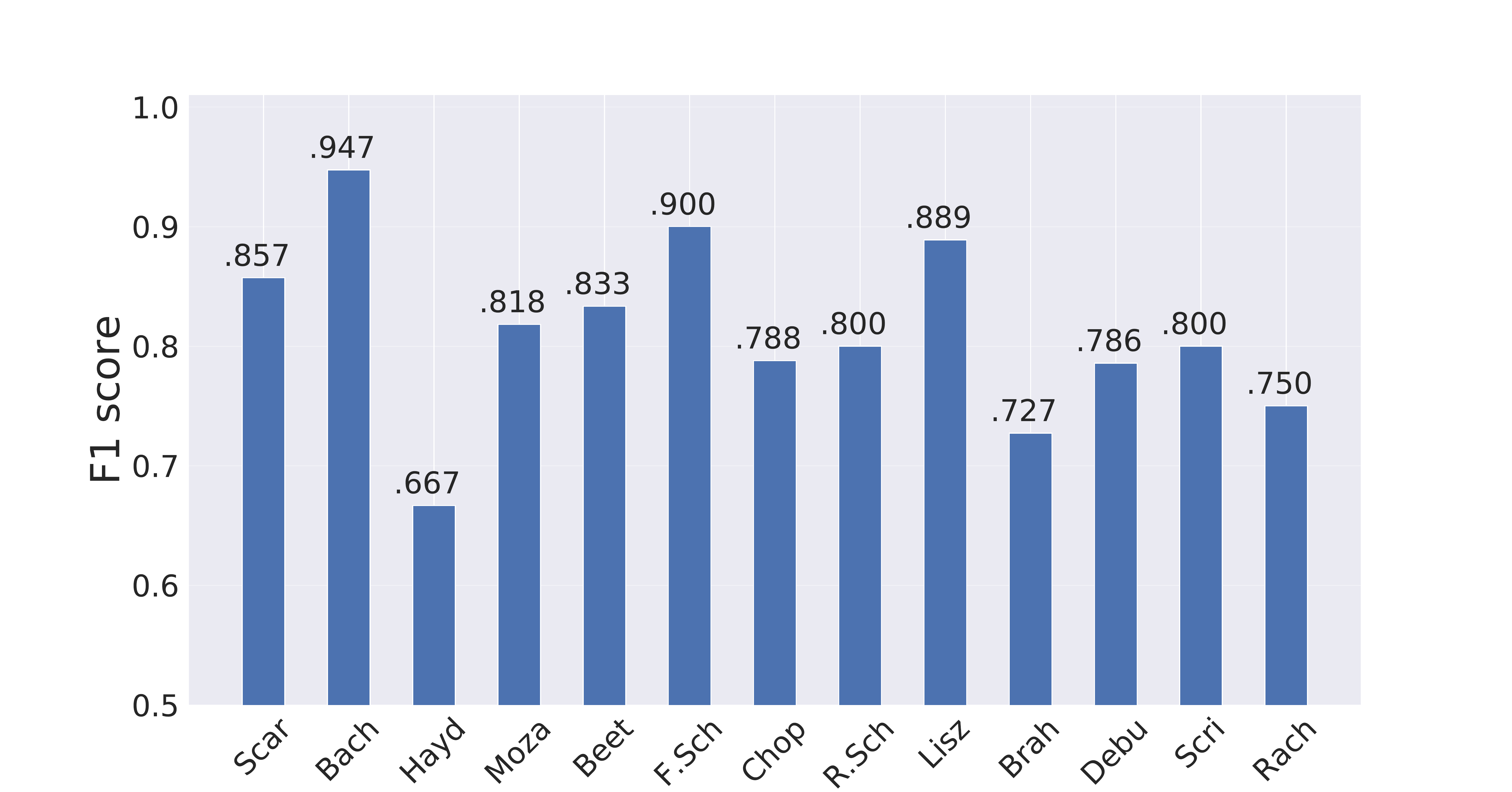}}}
 \label{fig:barchart}
 \caption{The F1 scores of each composer of the best performing model, sorted by the birth year of the composers.}
\end{figure}

After a preliminary experiment, we chose ResNet-50 as a reference model based on its performance after testing ResNet-18/34/50/101. The model was trained using SGD with momentum and weight decay. We used CE loss function with cosine annealing for LR scheduling. 

\subsection{Result and Discussion}
\vspace{-0.1cm}
We report our reference model with `weighted' averaged F1 score of $0.8333$.\footnote{ResNet-18/34/101 achieved F1 scores of $0.7962$/$0.7881$/$0.7892$.} 
This can be loosely compared to \cite{jainanalysis, CRNNcomposer} although the problem definition and the dataset vary. \cite{jainanalysis} using gradient boosting and CNN achieved F1 scores of $0.742$ and $0.700$ for classifying 15 and 6 classical composers, respectively. Similarly, a ConvRNN-based system in \cite{CRNNcomposer} achieved 70\% accuracy for classifying 6 composers. 

Based on this reference model,  
we investigated the performance variation by composers. First, the Spearman rank correlation coefficient of the composers' birth years and performances was $-0.45$, i.e., the model performed better for relatively old classical composers. This results seem musically plausible and related to the fundamental property of the problem -- Because the style of music has been diversified and more complicated chords have been popularized over time, recognition of composer by the music may be inherently more difficult for relatively modern composers. Second, we also hypothesized if the label imbalance in the training data has an effect on the performance, but the Spearman rank correlation coefficient between the number of training items (pieces) and performance was $-0.13$, which does not seem significant.

\begin{figure}
\small
 \centerline{\mbox{
 \includegraphics[width=0.95\columnwidth, clip]{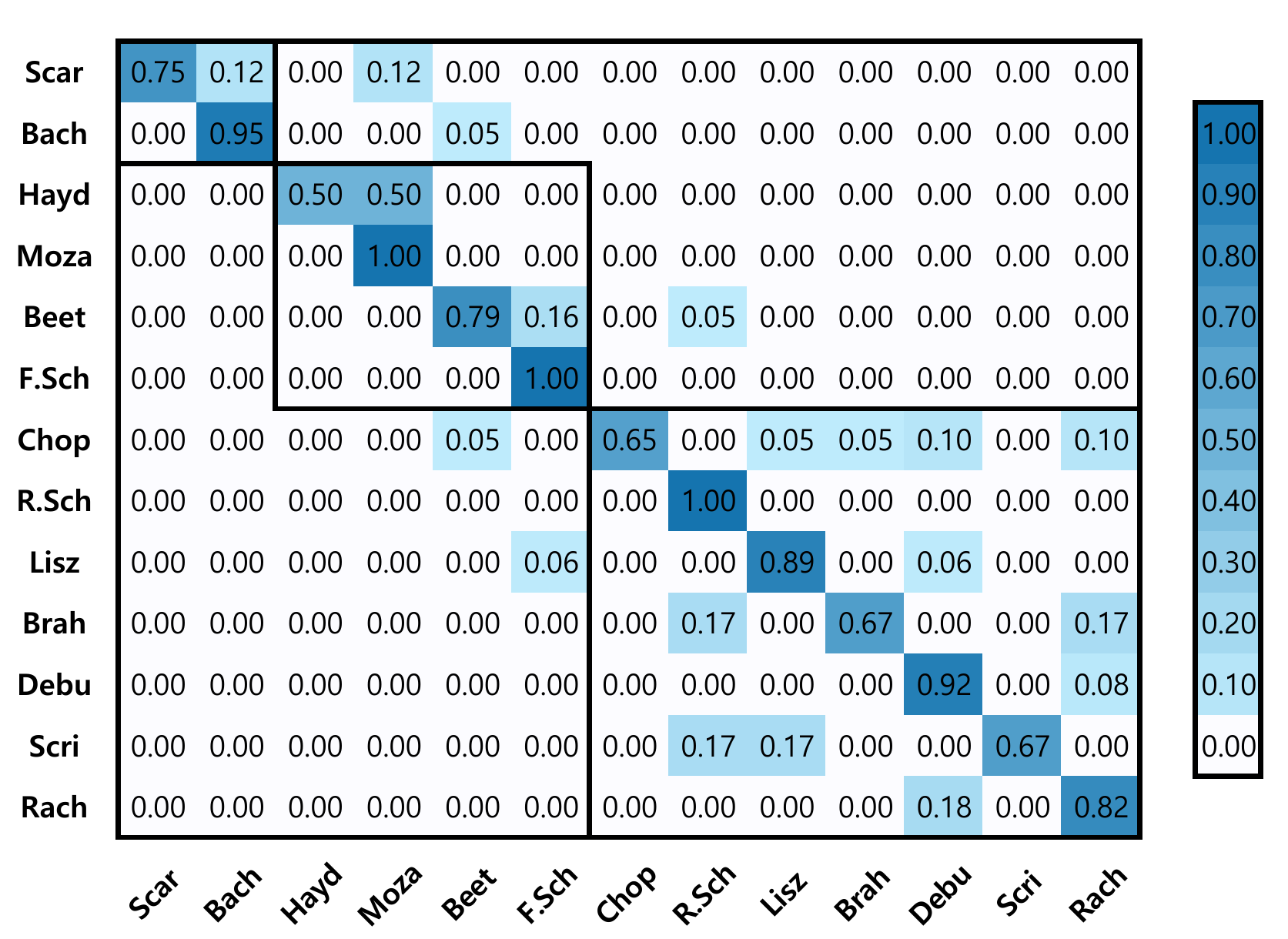}}}
 \caption{The confusion matrix of the classification result on the validation set sorted by the birth year. Bold boxes denote the musical eras.}
  \label{fig:confusion}
\end{figure}

The confusion matrix in Figure \ref{fig:confusion} shows that a majority of the misclassifications occurred within the same era (Denoted with bold boxes). Out of 19 misclassified pairs, only 5 were classified as composers from different era.
We assume the composers from the same era are more likely to share similar musical patterns, 
and this indicates that the model successfully learned such patterns during training.


\begin{table}
\small
 \begin{center}
\begin{tabular}{@{}c|c||c|c||c|c@{}}
    \multicolumn{2}{c||}{\textbf{No. of Segments}} &  \multicolumn{2}{|c||}{\textbf{Onset Channel}} &  \multicolumn{2}{|c}{\textbf{Frame Channel}} \\
\hline
\hline
5 & .5713 & Used & .8333  & Continuous & .8333\\
10 & .7196 & Omitted & .7858  &  Binarized & .8525\\
20 & .7687 & & & &\\
30 & .8148 & & & & \\
60 & .8249 & & & & \\
90 & .8333 & & & & \\

\end{tabular}
\end{center}
 \caption{Ablation study results (F1 score) of controlling i)~the number of segments, ii)~onset channel usage, and iii)~frame channel binarization.}
 \label{tab:results}
\end{table}


\vspace{-0.1cm}
\subsection{Ablation Study}




\textbf{Number of segments}: We trained various instances of the model by increasing the number of segments from 5 to 90 as reported in the first column of \tabref{tab:results}. The result indicates that segments over 30 provide information that is diverse enough for the model to perform at a certain level (over 80\% of F1 score). 

\noindent
\textbf{Onset Channel Usage}: As in the second column of \tabref{tab:results}, removing the onset channel from the input introduced a degradation of performance by $0.0475$. This result demonstrates two interesting aspects of the task and the model. First, even only with the frame channel, i.e., by comparing the chord and voicing, the model can classify composers with a reasonable performance. Second, the location of onsets may be helping to solve the task by clarifying the chords and/or providing rhythmic information. 


\noindent
\textbf{Frame Binarization}: In the third column of \tabref{tab:results}, we compare the result of our reference model (annotated as ``Continuous'') with a model that was trained using binarized frame channel, (i.e., one where the velocity values are binarized) in an attempt to maximize the symbolic characteristic of our data by removing the remaining performer-related information. 
Using this new input, somewhat surprisingly, this ``Binarized'' model achieved an F1 score of $0.8525$, outperforming the reference model by $0.0192$. 


\subsection{Conclusion}
In this paper, we showed that a composer classification can be done using symbolic representation. We investigated the trained model by varying the within-piece coverage and input data representation as well as analyzing the performances per composer. A future direction may include separating the influence of players, generalizing towards other instruments, and domain adaptation to other genres.

\bibliography{ISMIRtemplate}

\end{document}